\documentclass[prl,twocolumn,superscriptaddress]{revtex4-1} 
\usepackage[english]{babel}
\usepackage{amsmath, bm, amssymb}
\usepackage{mathtools}
\usepackage{siunitx}
\usepackage{environ}
\usepackage{mathrsfs}
\usepackage{braket}
\usepackage[]{hyperref}
\DeclareMathAlphabet{\mathscrlower}{OT1}{pzc}{m}{it} 
\usepackage{prettyref}
\usepackage{graphicx}
\usepackage{booktabs}
\usepackage{array}
\usepackage{multirow}
\usepackage{dcolumn}
\usepackage{threeparttable}
\usepackage{threeparttablex}
\usepackage{placeins}
\usepackage[stable]{footmisc}
\usepackage{mhchem}
\usepackage{xfrac}
\usepackage{upgreek}

\let\nablatmp\nabla
\renewcommand{\nabla}{\vec{\nablatmp}}
\DeclarePairedDelimiter\abs{\lvert}{\rvert}
\makeatletter
\let\oldabs\abs
\def\abs{\@ifstar{\oldabs}{\oldabs*}}

\AtBeginDocument{
\heavyrulewidth=.08em
\lightrulewidth=.05em
\cmidrulewidth=.03em
\belowrulesep=.65ex
\belowbottomsep=0pt
\aboverulesep=.4ex
\abovetopsep=0pt
\cmidrulesep=\doublerulesep
\cmidrulekern=.5em
\defaultaddspace=.5em
}
\begin{document}
\title{Stopping mass-selected alkaline-earth metal monofluoride beams of high energy via
formation of unusually stable anions}
\date{\today}
\author{Konstantin Gaul}
\affiliation{Fachbereich Chemie, Philipps-Universit\"{a}t Marburg, Hans-Meerwein-Stra\ss{}e 4, 35032 Marburg}
\author{Ronald F. Garcia Ruiz}
\affiliation{Massachusetts Institute of Technology,~Cambridge,~MA~02139,~USA}
\author{Robert Berger}
\affiliation{Fachbereich Chemie, Philipps-Universit\"{a}t Marburg, Hans-Meerwein-Stra\ss{}e 4, 35032 Marburg}
\begin{abstract}
Direct laser-coolability and a comparatively simple electronic
structure render alkaline-earth metal monofluoride molecules
versatile laboratories for precision tests of fundamental physics. In
this theoretical work, a route for efficient stopping and cooling of
high-energy hot beams of mass-selected alkaline-earth metal monofluorides via
their anions is explored to facilitate subsequent precision experiments with trapped molecules. It is shown that these molecular anions
possess an unusually strong chemical bond and that radium monofluoride anion, RaF$^-$, features 
properties favourable for decreasing kinetic and interal energy, indicating applicability
of direct laser-cooling of the anion.
\end{abstract}

\maketitle

\emph{Introduction.---}
Precision spectroscopy of cold molecules serves as a powerful tool for
studying fundamental symmetries \cite{demille:2015} and, therewith,
provides some of the most stringent tests of physics beyond the
Standard model of particle physics (BSM) \cite{roussy:2023}. Due to a
relatively simple electronic structure and the possibility for direct
laser-cooling, alkaline-earth metal monofluoride molecules are
particularly attractive for this purpose
\cite{isaev:2010,Isaev:2013,Isaev:13,barry:2012,zhelyazkova:2014,altundas:2018,aggarwal:2018,aggarwal:2021}.
As BSM effects are relativistic in nature, molecules containing heavy
elements such as radium monofluoride (RaF) appear as promising
candidates for fundamental physics studies \cite{isaev:2010}.  The theoretical
prediction of RaF as a versatile laser-coolable molecular probe of
fundamental physics
\cite{isaev:2010,Isaev:2013,Isaev:13,kudashov:2014,gaul:2017,gaul:2019,gaul:2020}
led recently to its spectroscopic identification and characterisation
at ISOLDE, CERN \cite{garciaruiz:2020}. This experiment paved the way
to study nuclear structure of various short-lived isotopes in molecular
systems in general and presented a first step in the effort to search
for violations of fundamental symmetries in RaF
\cite{udrescu:2021,udrescu:2024,wilkins:2023}. 

In, e.g. neutralisation set-ups as proposed
\cite{Isaev:13} and demonstrated \cite{garciaruiz:2020} for RaF, the
molecules emerge as a rather hot ensemble upon production: several
rovibrational levels are occupied and the molecules possess high
translational energy of several keV. In order to enable an efficient direct laser-cooling for subsequent precision experiments of trapped molecules with short-lived isotopes, the
molecular beam needs to be decelerated down to sub mK ($<0.1~\upmu$eV) energies. Such a great deceleration, however, is much
more efficient, if those beams consist of ions. The cationic species
of alkaline-earth metal monofluorides have a similar electronic structure as
RaF$^+$ and nearly parallel potential energy curves to the neutrals in
their electronic ground state \cite{isaev:2010,garciaruiz:2020}, which
is favourable for efficient neutralisation with little change in
vibrational quantum numbers. However, the neutralisation of \ce{RaF+}
in collisions with other elements can be expected to significantly
increase the internal energy of the molecule. 

\begin{figure}[!hbt]
\includegraphics[width=.5\textwidth]{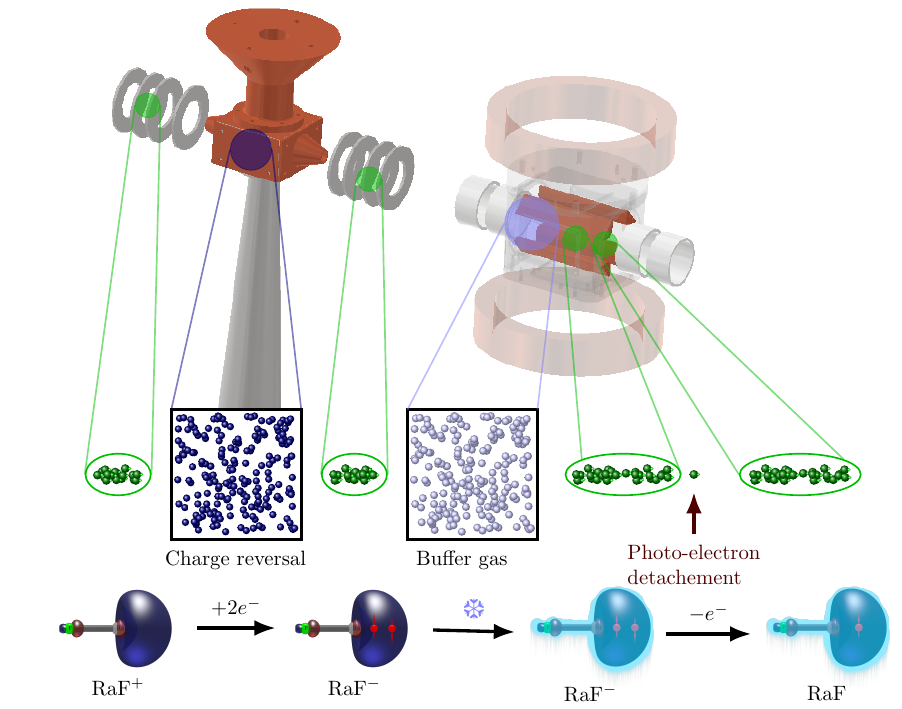}
\caption{Schematic of the proposed set-up to create a slow, cold beam
of RaF from a source of \ce{RaF+} molecules by charge reversal to
\ce{RaF-}, which is subsequently stopped and cooled in a gas-filled
radio frequency Paul trap, and, finally, neutralised via photo-electron detachment with
a laser within the ion trap.}
\label{fig: setup}
\end{figure}

An alternative way to decelerate and cool a neutral molecular beam
would be to use the anionic species as a precursor
\cite{hamamda:2015a}, which can be stopped and cooled in a gas-filled
radio-frequency ion trap at cryogenic temperature as sketched in
Figure \ref{fig: setup}. In contrast to a cation, an anion can be
neutralised comparatively easily by photo-electron detachment with a
laser afterwards, without unwanted excitation or heating. For this
purpose it can be beneficial if the
dissociation energy $D_\mathrm{e}$ of the anion lies well above its
adiabatic electron-detachment energy (ADE) or wavenumber (ADW). The
concept discussed in Ref.~\cite{hamamda:2015a} refers, however, to
dipole-bound anions: i.e. highly polar closed-shell neutral molecules to which
an additional electron is loosely bound by only several meV. In the open-shell alkaline-earth metal
monofluorides, in contrast, the additional occupation of a valence orbital with concomitant spin-pairing upon anion formation has to be taken
into account. Moreover, in Ref.~\cite{hamamda:2015a} dissociative
electron attachment was not considered. From naive chemical intuition
based on filled subshells, the bonding in alkaline-earth metal monofluoride
anions would be expected to be very weak: Two fragments with closed
electronic subshells would exhibit strong Pauli repulsion for short
internuclear distances due to overlapping occupied orbitals and, in
the absence of covalent interaction, show mostly weakly attractive
dispersive interactions for larger distances. Opposed to this
expectations, it was reported for \ce{BeF-}, however, that a very stable
dative bond is formed between Be and \ce{F-}, which was attributed to
the strong Lewis acidity of Be \cite{green:2018}.

In this letter we show with accurate coupled cluster (CC) calculations
that the charge reversal of \ce{MF+} to \ce{MF-} and subsequent
photo-electron detachment to neutral MF is expected to be feasible for
all alkaline-earth metal monofluorides and does not emerge as a special
property of Be. This route can provide access to stopped cold samples of
mass-selected neutral \ce{MF} with short-lived isotopes.
We provide an understanding of the unprecedented
stability of \ce{MF-} by analysing chemical bonding within symmetry
adapted perturbation theory (SAPT) and population analysis. Moreover,
we study electronic excitation in \ce{RaF-} and estimate branching
fractions via Franck--Condon factors for several relevant vibronic
transitions in \ce{RaF-} and from \ce{RaF-} to RaF to assess the
efficiency of this pre-cooling scheme. Our study provides also a hint
for the applicability of our scheme in other diatomic or small
polyatomic molecules with a similar electronic structure
\cite{isaev:2016}. Additionally, we identify properties that are
required to achieve direct laser-coolability of molecular anions. 

\begin{table}
\caption{Dissociation wavenumbers $\tilde{D}_\mathrm{e}$, adiabatic 
photo-electron detachment wavenumbers (ADW), equilibrium bond length $r_\mathrm{e}$ and harmonic
vibrational wavenumber $\tilde{\omega}_\mathrm{e}$ of \ce{MF-}
computed at the level of RECP-CCSD(T)}
\begin{tabular}{lS[round-mode=figures,round-precision=3]S[round-mode=figures,round-precision=3]S[round-mode=figures,round-precision=3]S[round-mode=figures,round-precision=3]}
\toprule
Molecule & {$\tilde{D}_\mathrm{e}/\si{\per\centi\meter}$} &
{$\mathrm{ADW}/\si{\per\centi\meter}$} & 
{$r_\mathrm{e}/\si{\angstrom}$} &
{$\tilde{\omega}_\mathrm{e}/\si{\per\centi\meter}$} \\
\midrule
\ce{BeF-} & 28672.88 & 8586.69 & 1.4153 &1062.83  \\
\ce{MgF-} & 21592.97 &10799.65 & 1.8181 & 577.48 \\
\ce{CaF-} & 26202.62 & 8295.45 & 2.0118 & 515.94  \\
\ce{SrF-} & 24997.95 & 7946.19 & 2.1508 & 434.29 \\
\ce{BaF-} & 26842.36 & 6774.20 & 2.2671 & 439.36  \\
\ce{RaF-} & 23617.00 & 7214.88 & 2.3596 & 400.96   \\
\bottomrule
\end{tabular}
\label{tab: results_other}
\end{table}

\begin{figure}
\includegraphics[width=.5\textwidth]{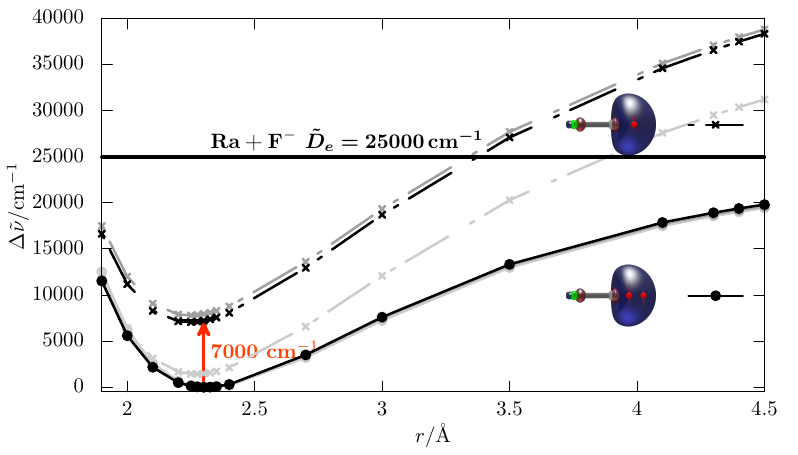}
\caption{Potential energy curves of \ce{RaF} and \ce{RaF-} computed at
the level of 4c-CCSD(T) (black line), 4c-CCSD (dark gray line) and
4c-HF (light gray line). The potential curves of \ce{RaF} are shown with
broken lines and crosses, whereas the potential curves of \ce{RaF-} are shown with
solid lines and dots. For the anion the potential curves on different levels of
theory (4c-HF, 4c-CCSD, 4c-CCSD(T)) overlap each other, such that only
the 4c-CCSD(T) potential is visible. The photo-electron detachment
energy is indicated by a transition arrow from lowest vibrational
state of \ce{RaF-} to the lowest vibrational state of \ce{RaF}. The
given dissociation energy is computed by separate calculations of
\ce{RaF-}, Ra and \ce{F-}. Lines between individual points are shown
to guide the eye.}
\label{fig: ade_diss}
\end{figure}

\emph{Stability of alkaline-earth metal monofluoride anions.---}
One of the most important criteria for the use of molecular anions as 
precursors of cool neutral molecules is their stability with respect to
dissociation. We determined the bond dissociation energies of \ce{MF-},
with M denoting here and in the following a group 2 element,
at the coupled cluster level of theory including singles and doubles
amplitudes and accounted perturbatively for triples amplitudes
[CCSD(T)] and accounted for scalar relativistic effects with effective core 
potentials (RECP). Further computational details are provided in the 
Supplemental Material. Tab.~\ref{tab: results_other} lists the results
obtained. We find $\tilde{D}_\mathrm{e}
\gg \mathrm{ADW}$ for all \ce{MF-}, which indicates that using anions
to generate cold samples of neutral alkaline-earth metal monofluorides appears 
to be favorable for all alkaline-earth metals. Our results for \ce{BeF-} are
in very good agreement with experimental values
($\mathrm{ADW}=\SI{8697(6)}{\per\centi\meter}$,
$\tilde{\omega}_\mathrm{e}=\SI{1059(6)}{\per\centi\meter}$) and coupled cluster
calculations of Ref. \cite{green:2018}. For the particular case of
\ce{RaF-}, we studied the influence of relativistic effects on the
Dirac--Coulomb four-component coupled cluster level [4c-CCSD(T)] and
considered vibrational effects as well. The results from different
methods for \ce{RaF-} are compared in the Supplemental Material.
We compute a dissociation energy of \ce{RaF-}
with
$\tilde{D}_\mathrm{e}=\SI{25000(1200)}{\per\centi\meter}hc=\SI{3.10(16)}{\electronvolt}$
and find that its adiabatic electron detachment energy energy lies far below at
$\mathrm{ADW}=\SI{7060(640)}{\per\centi\meter}=\SI{0.875(79)}{\electronvolt}$,
where we estimate the given error bars as detailed in the supplemental
material.
Vibrational corrections to the ADW are on the order of
$\SI{28}{\per\centi\meter}$ for the 0-0 transition and, therefore, 
negligible. These results are summarised in Figure \ref{fig:
ade_diss}. The laser needed for resonant photo-electron detachment
from \ce{RaF-} would be operating in the near infrared at about
\SI{1400}{\nano\meter}, which would be feasible with standard
laser technology.

\emph{Chemical bonding in RaF$^-$ and lighter homologues.}---
The dissociation energies show that the additional electron, which
occupies a molecular orbital of $\sigma^*$ type, leads to a surprisingly small
destabilisation of the chemical bond of \ce{MF-} molecules. This is
attributed to the essentially non-bonding character of the
corresponding orbital. Bonds become elongated and harmonic vibrational
wavenumbers reduced, but only slightly so from MF to \ce{MF-}. Bond
formation between a neutral alkaline-earth metal atom and a fluoride
anion would naively be expected to be very weak as the fully filled
subshells of M and \ce{F-} should strongly repel each other. For
\ce{BeF-} it was shown, however, that this anion forms a very stable 
bond, which was argued to be a special property of Be and thus not 
to be expected in other group 2 fluoride anions \cite{green:2018}. We have
demonstrated above that it is not a specific feature of \ce{BeF-}, but
appears instead for all \ce{MF-}. To investigate the special bonding situation
in \ce{MF-} further and complement the natural bond analysis done in
Ref. \cite{green:2018}, we performed an SAPT-based energy
decomposition analysis (EDA) \cite{jeziorski:1994} for all alkaline-earth metal monofluorides from
\ce{BeF-} to \ce{RaF-} and compared to the cations \ce{BeF+} to
\ce{RaF+}, which are viewed to be strongly bound due to
electro-static interactions of \ce{F-} and \ce{Ra^2+}. We performed an
SAPT to second order, decomposing the bonding energy as
\begin{align}
D_\mathrm{e} &= \Delta E_\mathrm{prep} + \Delta E_\mathrm{int}\\ 
\Delta E_\mathrm{int} &= \Delta E_\mathrm{elstat} + \Delta
E_\mathrm{ex} + \Delta E_\mathrm{ind} +\Delta E_\mathrm{dsp}+ \Delta E_\mathrm{HF}\,.
\end{align}
SAPT contributions to the bonding interaction energy
$\Delta{}E_\mathrm{int}$ are split up into first order electrostatic
interactions $\Delta{}E_\mathrm{elstat}$, first order Pauli repulsion
$\Delta{}E_\mathrm{ex}$ due to overlap of charge densities, second
order polarisation effects due to induction from the metal atom to
fluoride $\Delta{}E_\mathrm{ind,M\rightarrow{}F}$, from the fluoride
to the metal atom $\Delta{}E_\mathrm{ind,F\rightarrow{}M}$, due to
dispersive interactions $\Delta{}E_\mathrm{dsp}$ and higher order
effects estimated as difference to the Hartree--Fock (HF) bonding
energy $\Delta{}E_\mathrm{HF}$. $\Delta E_\mathrm{prep}$ is the energy
needed to prepare the atomic fragments to the bonding state.  SAPT
data is visualised in Figure \ref{fig: eda} and all numerical results
of SAPT and partial charges are provided in the Supplementary
Material.
For interpretation of the SAPT-EDA scheme used herein, we have to note
that induction terms contain not only polarisation but also
charge-transfer contributions \cite{misquitta:2013}. As of that,
induction effects can be a measure for charge-transfer if polarisation
effects are small \cite{misquitta:2013}. The
arrows above indicate the direction of polarisation, this means
$E_{\mathrm{ind},\ce{F}\rightarrow\ce{M}}$ contains a polarisation of
M by F, and the arrows point also in the direction of a possible flow
of charge, i.e.  $E_{\mathrm{ind},\ce{F}\rightarrow\ce{M}}$ contains
transfer of \emph{positive} charge from \ce{F} to \ce{M} (or transfer
of electrons in the opposite direction). Thus,
$E_{\mathrm{ind},\ce{M}\rightarrow\ce{F}}$ implies electron-transfer
from \ce{F} to \ce{M} and polarisation of F by M. Moreover, ECP
calculations tend to underestimate the exchange induction contribution
\cite{patkowski:2007} leading to an underestimation of the induction
energy portion. To analyse this ECP artefact, we computed \ce{CaF+}
and \ce{CaF-} both with and without an ECP. Accordingly, the ECP
causes a change in the induction term
$E_{\mathrm{ind},\ce{M}\rightarrow\ce{F}}$ of \ce{CaF-} by
\SI{-1.3}{\electronvolt} (see Figure \ref{fig: eda}). When assuming a similar impact on
$E_{\mathrm{ind},\ce{M}\rightarrow\ce{F}}$ for heavier homologues, we
find that $E_{\mathrm{ind},\ce{M}\rightarrow\ce{F}}$ amounts between
\SI{-0.8}{\electronvolt} and \SI{-1.3}{\electronvolt} for \ce{SrF-} to
\ce{RaF-}. The observed trend of $E_{\mathrm{ind},\ce{M}\rightarrow\ce{F}}$
for \ce{CaF-} to \ce{RaF-} parallels the trend of positive electron
affinities of Ca, Sr, Ba and Ra reported in Ref.~\cite{andersen:1997}.

As the HF dissociation energies deviate only mildly from CCSD(T)
values, we assume that correlation effects will not have a pronounced
influence on the SAPT-based analysis. This energy analysis is
accompanied by a simple Mulliken population analysis to show the
charge distribution in the bonded molecules. Understanding of
chemical bonding situations is generally important for specific design
of molecules in fundamental physics applications, as shown earlier for
instance by successful identification of polyatomic candidate
molecules for laser-cooling \cite{isaev:2016}.

\begin{figure}
\includegraphics[width=.5\textwidth]{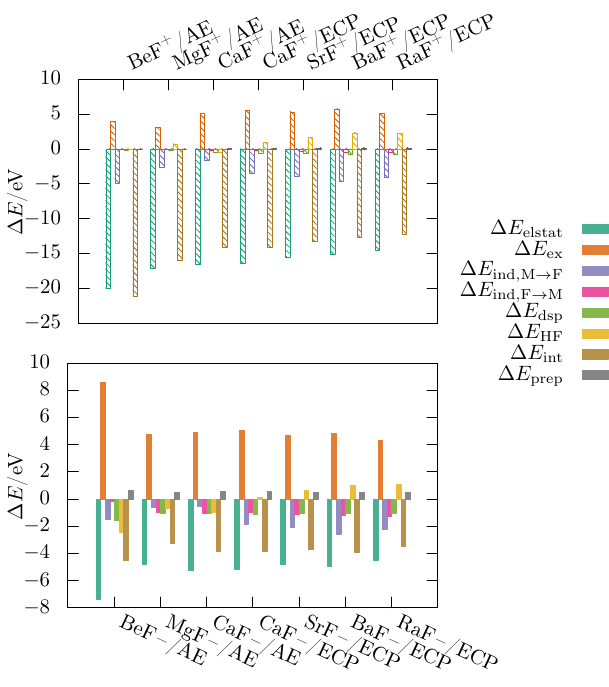}
\caption{Visualisation of energy contributions to the total bonding in
alkaline-earth metal monofluoride anions (\ce{MF-}) and cations
(\ce{MF+}) within SAPT as described in the text.}
\label{fig: eda}
\end{figure}

Mulliken partial charges in our \ce{MF} ions resemble essentially the naive picture of a
charge distribution with \ce{M^2+} and \ce{F-} in \ce{MF+}, and \ce{M}
and \ce{F-} in \ce{MF-}, in particular for the alkaline-earth metals
heavier than Mg. According to SAPT, the strong bond in \ce{MF+} results mainly from 
strong electrostatic interactions, whereas in \ce{MF-} electrostatic
interactions are, as one would likely expect, about the same size as the Pauli repulsion of the
overlapping electron clouds of \ce{Ra} and \ce{F-}. Bonding in the anion
appears in this picture essentially only in second order due to inductive and
dispersive effects.

Stabilising induction effects of \ce{F} by \ce{M},
$E_{\mathrm{ind},\ce{M}\rightarrow\ce{F}}$, are pronouncedly larger in
magnitude in the cations than in the anions. This observation agrees
with a much larger energy gain by partial electron transfer from
\ce{F-} to \ce{M^2+} in \ce{MF+} than by an electron transfer from
\ce{F-} to \ce{M} in \ce{MF-}, naively expected from an increased
electron affinity of the dication.  The induction in the opposite
direction $E_{\mathrm{ind},\ce{F}\rightarrow\ce{M}}$ yields an equally
important stabilising contribution to the bonding in the heavier
homologues for \ce{MF-}, whereas this contribution is negligible for
\ce{MF+}.  Moreover, such an inductive effect is also absent in weakly
bound systems, in which dispersion forces dominate such as the
Ar dimer \cite{patkowski:2007}.  The
pronounced contributions from induction of M by F in the other anions
along with contributions of dispersion and higher order effects,
however, indicate a more complex bonding mechanism than described by a
dative \ce{F$\rightarrow$M}-bond proposed for \ce{BeF-} in Ref.
\cite{green:2018}. From the SAPT-EDA considerable backdonation from the
metal atom to the fluoride anion can be expected. Our general findings of a strong
chemical bond in \ce{MF-} were subsequently also confirmed by a
different approach in Ref.~\cite{liu:2023}, in which \ce{BeF-} to
\ce{BaF-} were considered.

\emph{Experimental feasibility to create cold neutral molecules from a beam
of anionic precursors.}---
In the following, we analyse the experimental feasibility to create cold
samples of \ce{MF} from \ce{MF-} beams on the specific
example of \ce{RaF-}. Franck--Condon factors for the transition from
\ce{RaF-} to \ce{RaF} indicate that the branching fraction for occupying
one of the four lowest vibrational states in RaF approaches $99.9\,\%$ if the
electron is detached with high excess energy from \ce{RaF-} in its
vibrational ground state. This renders direct vibrational pre-cooling of \ce{RaF-} 
particularly interesting, as resulting neutral RaF molecules would end up mainly in the
lowest vibrational states upon photo-electron detachment, whereas upon
threshold photo-electron detachment, only the lowest vibrational state
of the neutral would be occupied with a still favourable Franck--Condon
factor. 

Moreover, we investigate possibilities for direct laser-cooling of
\ce{RaF-}. For this purpose we determined the four energetically
lowest lying electronically excited states of \ce{RaF-} using an
Equation-of-Motion 4c-CCSD approach (EOM-4c-CCSD). The results are
shown in Fig.~\ref{fig: excitation} and further information is
provided in the Supplemental Material.  We assume the theoretical
uncertainty of excitation energies at the level of EOM-4c-CCSD for the
basis set used to be on the order of $\sim20\%$ as described in the
Supplementary Material.

We find that the lowest two electronic states lie less than
\SI{1600}{\per\centi\meter} below the ADW (see Fig.~\ref{fig:
excitation}). This closeness in energy might render optical cycling in 
\ce{RaF-} more difficult. Apart from that, our calculations suggest that the
bonding situation in \ce{RaF-} is not affected much by excitation in these
electronic states, so that we find rather parallel potential energy curves. Harmonic
vibrational wavenumbers are slightly increased and differ by less
than $8\,\%$ from the one of the electronic ground state and bonds are
slightly elongated by less than $2\,\%$. Both indicate a slight
stabilisation of the chemical bond upon electronic excitation. 
Electronic excitation energies for the two lowest states, which have
$\Omega=0$, remain below the electron detachment energy and are
almost degenerate at $\tilde{T}_\mathrm{e}=\SI{6300(1400)}{\per\centi\meter}$
with a predicted splitting of only about $\SI{90}{\per\centi\meter}$. 

\begin{figure}
\includegraphics[width=.5\textwidth]{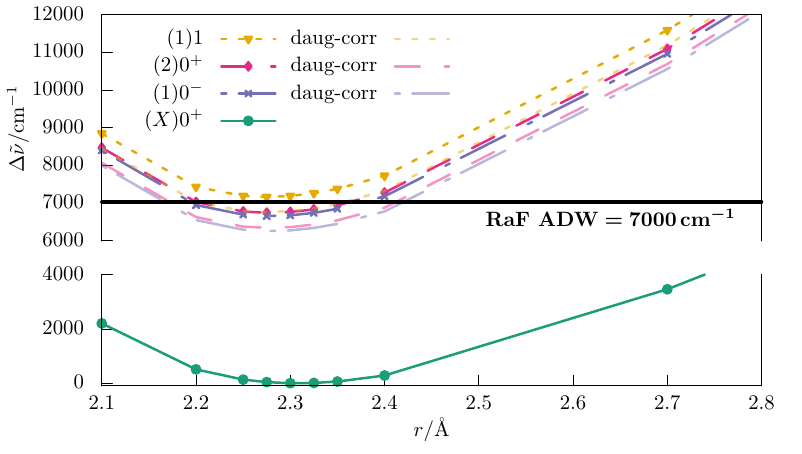}
\caption{Potential energy curves of the electronic ground state and
the four energetically lowest electronically excited states of
\ce{RaF-} computed at the level of EOM-4c-CCSD. The given dissociation
energy is computed by separate calculations of \ce{RaF-}, Ra and
\ce{F-} at the level of 4c-CCSD(T)/dyall.cv4z. The effect of double
augmentation of the basis set was calculated at the equilibrium
distance only and is indicated as a constant shift of the potential by
transparent lines of the same colour and dash type.  Lines between
individual points are shown to guide the eye.}
\label{fig: excitation}
\end{figure}

To estimate the oscillator strength of these transitions, we computed
electronic transition matrix elements on the level of relativistic
time-dependent density functional theory (TDDFT). The TDDFT-CAMB3LYP
calculations reproduce the EOM-4c-CCSD excitation energies reasonably
well (see Supplementary Material).  All electronic states below or
slightly above the ADE can be assumed to be of triplet character and,
thus, transitions from the electronic ground state show very low
oscillator strengths.  We find that one-photon transition to the
lowest excited state is forbidden within the numerical precision.
Although the oscillator strength for the second excited state is
rather low as well, and as of that corresponds to unfavourably long
lifetimes of $\sim\SI{100}{\micro\second}$, this excited state may 
still be amenable to direct laser-cooling of \ce{RaF-} as longer
interaction times can be achieved by using an ion trap
\cite{nguyen:2011,ivanov:2020}.

From EOM-4c-CCSD potentials, we receive cumulative Franck--Condon
factors of $>0.999$ for the $0^+$-state (second excited state),
when taking into account the vibrational transitions 0-0, 0-1 and 0-2.
Thus, this excitation could be interesting for direct laser-cooling as
the ADW could be expected to lie energetically above the vibrational
ground state of the excited $0^+$ state. Furthermore, we can conclude that
the efficiency of the pre-cooling scheme is not compromised either, as a
spread over many vibronic levels is not expected. 

Optionally, a hybrid trap consisting of a Paul trap and a magnetic
trap could be implemented to achieve efficient trapping and cooling
of the neutral RaF molecules \cite{mangeng:2023}.  Magnetic traps with
potentials as deep as a few Kelvin could be used to trap the neutral
molecules for further laser cooling or sympathetic cooling using
neutral atoms with simple electronic structure.

Finally in this subsection, we discuss a possible destruction of \ce{RaF-} by electron
detachment in collision with buffer gas atoms or by neutralisation due to
strong electrical fields. The first can be suppressed by using light
rare gases like Ne or He as buffer medium \cite{zappa:2001}, as these
have no or negligible electron affinities and comparatively low cross
sections for converting kinetic energy of the anion into internal
energy, so that these collision partners are expected to account for
low neutralisation rates. To estimate neutralisation due to an
electric field, we can employ a very simple model from
Ref.~\cite{schweinler:1980} in which the neutralisation rate $w$ of an
anion in a spherical electronic state is given as a function of the
field strength $\mathcal{E}_z$ in units of
$\si{\hartree\per\elementarycharge\per\bohr}\approx\SI{5e9}{\volt\per\centi\meter}$,
ADE in units of $\si{\hartree}$, the size of the anion
$r_0$ in units of $a_0$ and the fraction of charge density outside
this sphere $f_0$ as
\begin{multline}
w = \SI{2e16}{\hertz}\\
\times\frac{\mathcal{E}_z a_0 e}{E_\mathrm{h}}\frac{
f_0\mathrm{exp}\{2\sqrt{\frac{2\mathrm{ADE}}{E_\mathrm{h}}}\frac{r_0}{a_0}-2\frac{(2\mathrm{ADE}/E_\mathrm{h})^{3/2}}{3e
a_0\mathcal{E}_z/E_\mathrm{h}}\}}{\sqrt{2\mathrm{ADE}/E_\mathrm{h}}} 
\end{multline}
Assuming that 99~\% of the charge of \ce{RaF-} is contained in a
sphere of radius $r_0=2r_\mathrm{e}$ the neutralisation rate would be
well below \SI{1}{\milli\hertz} for fields of strength
$<\SI{1e6}{\volt\per\centi\meter}$. This can be neglected for the desired experimental conditions.

\emph{Production of \ce{RaF-}.---}
If not specifically produced in negative ion sources (see for instance
Ref.~\onlinecite{alton:2005} for a review), \ce{RaF-} can be formed via
collisional charge reversal from its cationic precursor as illustrated
in Figure \ref{fig: setup}. The first
ionisation energy of RaF is approximately \SI{4.9}{\electronvolt}
\cite{Isaev:13,garciaruiz:2020} and, thus, the total energy required to
generate \ce{RaF-} out of a \ce{RaF+} beam is about
$\sim\SI{5.8}{\electronvolt}$. Collisional charge reversal (or charge
inversion) from molecular cations to anions, established in mass
spectrometry since the 1970s (see
Refs.~\onlinecite{keough:1973,hayakawa:2001,hayakawa:2004,he:2004}), is
expected to be possible with conversion efficiencies larger than $5\,\%$
\cite{keough:1973,schlachter:1980,hayakawa:2004} with alkali atom
vapours. Creating the anion by collisions with Rydberg
atoms \cite{desfrancois:1994} as discussed for dipole-bound anions in
Ref.~\cite{hamamda:2015a} is expected to be inefficient due to the
comparatively large electron affinity of \ce{RaF-}. Mechanisms
discussed for collisional anionisation of cationic precursors include
single-collision double electron transfer with concomitant linear
dependence on the vapour pressure as well as sequential single electron
transfer that displays quadratic pressure dependence
\cite{keough:1973,hayakawa:2001,hayakawa:2004,he:2004}. Whereas
single-collision double electron transfer would be highly non-resonant by
virtue of large ionisation energies of alkali metal monocations $A^+$ ($>
\SI{22}{eV}$), the first neutralisation step of sequential electron transfer to
\ce{RaF+} proceeds nearly resonantly \cite{Isaev:13} and highly efficiently with Na
\cite{garciaruiz:2020}, while the second electron transfer step is
endothermic by about 4.2~eV with sodium vapours and by approximately 3~eV
with cesium vapours, suggesting that electron transfer is efficient for
fast neutrals with keV kinetic energies. Due to the special
electronic structures of alkaline-earth metal fluorides, dissociation
energies are high whereas structural changes in the three relevant charge
states are small, so that both double and single electron transfer are not
expected to induce pronounced fragmentation. By the same token, these
specific chemical features of these molecules will suppress isobaric
interferences from other isotopologues from other groups of the periodic
system of elements, thereby affording pure beams from mass-selected ionic
precursors.

\emph{Conclusion.---}
We proposed an efficient scheme for stopping mass-selected molecular
beams that can be employed to realise high-precision
neutralisation-reionisation spectroscopy experiments of short-lived
isotopologues of alkaline-earth metal monofluoride molecules which can
offer highly sensitive tests of BSM physics. We demonstrated that
alkaline-earth metal monofluoride anions possess an unusually stable
bond due to a complex bonding mechanism that contradicts naive
electrostatic bonding pictures. Moreover we showed with highly
accurate ab initio calculations exemplary for RaF that molecular beams can
be efficiently decelerated and trapped by formation of \ce{RaF-}, buffer-gas cooling
and successive photo-electron detachment without considerable
reheating. For all alkaline-earth metal monofluoride molecules, pre-cooling
via anions is expected to be feasible: the dissociation energy of
all \ce{MF-} lies well above its electron detachment energy. 

Finally, \ce{RaF-} has some favourable characteristics such as
highly-diagonal Franck--Condon factors of transitions to low-lying
electronically excited states that make it a compelling candidate to
explore direct laser-cooling of the anion, if remaining challenges
caused by low transition rates and an energetically closely lying
electron detachment threshold can be controlled. Laser cooling of
molecular ions is an open problem of interest in diverse fields, with
an impact on the study of antimatter \cite{yzombard:2015}.

Our study opens up new possibilities to generate ultra-cold
mass-selected samples of alkaline-earth metal monofluoride molecules, which
is of relevance for upcoming experiments with radioactive molecules
\cite{arrowsmithkron:2023}. In particular for RaF, the proposed scheme
will help to improve future precision studies of this molecule and
advance experiments \cite{garciaruiz:2020,udrescu:2021,udrescu:2024}
further towards detection of a violation of fundamental symmetries
\cite{isaev:2010} and to probe for new physics. The proposed scheme
may be transferable to other related species, including polyatomic
molecules \cite{isaev:2016}.

\emph{Acknowledgement.---}
Computer time provided by the NHR Center NHR@{}SW at Goethe-University
Frankfurt is gratefully acknowledged. This is funded by the Federal Ministry of
Education and Research, and the state governments participating on the basis of
the resolutions of the GWK for national high performance computing at
universities (www.nhr-verein.de/unsere-partner). We would like to thank A.
Vernon for the design of the charge exchange cell and ion trap. We thank Shane
G. Wilkins and other colleagues who attended the ISOLDE user meeting and
workshop 2021 or the 58th Symposium on Theoretical Chemistry in Heidelberg 2022
for discussions of our results. This work is funded by the Deutsche
Forschungsgemeinschaft (DFG, German Research Foundation) --- Projektnummer
445296313 as well as by DOE grants DE-SC0021176 and DE-SC0021179.

\bibliography{AK.bib}

\end{document}